\documentclass{sig-alternate-05-2015}

\usepackage{balance}  % to better equalize the last page
\usepackage{graphics} % for EPS, load graphicx instead
\usepackage{times}    % comment if you want LaTeX's default font
\usepackage{url}      % llt: nicely formatted URLs
\usepackage{xspace}
\usepackage{graphicx} % for using png as image format
\usepackage{float}
\usepackage{placeins}
\usepackage{dcolumn}
\usepackage{ulem}

\usepackage{color}

\def\sharedaffiliation{%  % from suggestion for shared affiliation from http://tex.stackexchange.com/questions/271627/shared-author-affiliation-in-acm-sig-proc
\end{tabular}
\begin{tabular}{c}}

\begin{document}

% Copyright
%\setcopyright{acmcopyright}
%\setcopyright{acmlicensed}
%\setcopyright{rightsretained}
%\setcopyright{usgov}
%\setcopyright{usgovmixed}
%\setcopyright{cagov}
%\setcopyright{cagovmixed}

% DOI
\doi{10.475/123_4}

% ISBN
\isbn{123-4567-24-567/08/06}

%Conference
\conferenceinfo{SWDM2016}{October 28, 2016, Indianapolis, USA}

%\acmPrice{\$15.00}

%
% --- Author Metadata here ---
\conferenceinfo{SWDM2016}{October 28, 2016, Indianapolis, USA}
%\CopyrightYear{2007} % Allows default copyright year (20XX) to be over-ridden - IF NEED BE.
%\crdata{0-12345-67-8/90/01}  % Allows default copyright data (0-89791-88-6/97/05) to be over-ridden - IF NEED BE.
% --- End of Author Metadata ---

\title{Project ACT: Social Media Analytics in Disaster Response}

%\subtitle{[Extended Abstract]
%\titlenote{A full version of this paper is available as
%\textit{Author's Guide to Preparing ACM SIG Proceedings Using
%\LaTeX$2_\epsilon$\ and BibTeX} at

\numberofauthors{1}
% Three authors sharing the same affiliation.
    \author{
    \alignauthor Wanita Sherchan, Shaila Pervin, Christopher J. Butler, Jennifer C. Lai \ \\
      \email{ \{wanita.sherchan, shaila.pervin, chris.butler\}@au1.ibm.com, jlai@us.ibm.com}
      \sharedaffiliation
      \affaddr{ IBM Research - Australia}\\
%      \affaddr{Victoria, Australia }
          }

%}
% There's nothing stopping you putting the seventh, eighth, etc.
% author on the opening page (as the 'third row') but we ask,
% for aesthetic reasons that you place these 'additional authors'
% in the \additional authors block, viz.
%\additionalauthors{Additional authors: John Smith (The Th{\o}rv{\"a}ld Group,
%email: {\texttt{jsmith@affiliation.org}}) and Julius P.~Kumquat
%(The Kumquat Consortium, email: {\texttt{jpkumquat@consortium.net}}).}
% Just remember to make sure that the TOTAL number of authors
% is the number that will appear on the first page PLUS the
% number that will appear in the \additionalauthors section.

\maketitle

\begin{abstract}
In large-scale emergencies social media has become a key source of information for public awareness, government authorities and relief agencies. 
However, the sheer volume of data and the low signal-to-noise ratio limit the effectiveness and the efficiency of using social media as an intelligence resource. 
We describe \textit{Australian Crisis Tracker} (ACT), a tool designed for agencies responding to large-scale emergency events, to facilitate the understanding of critical information in Twitter. 
%ACT uses the Twitter streaming API and processes each tweet through a pipeline of components- filters, metadata parser, and clusterer in order to group only the relevant tweets into events. Each of these events is then geocoded, categorised and augmented with images from Instagram. This pipeline of analytics is coupled to an intuitive and interactive  web interface that allows stakeholders to better access information during natural disasters. 
ACT was piloted by the \textit{Australian Red Cross} (ARC) during the $2013$-$2014$ Australian bushfires season. 
Video is available at: 

\noindent \textbf{\textit{https://www.youtube.com/watch?v=Y-1rtNFqQbE}}
\end{abstract}

%\begin{CCSXML}
%<ccs2012>
%  <concept_id>10010520.10010553.10010562</concept_id>
%  <concept_desc>Computer systems organization~Embedded systems</concept_desc>
%  <concept_significance>500</concept_significance>
%  <concept_id>10010520.10010575.10010755</concept_id>
%  <concept_desc>Computer systems organization~Redundancy</concept_desc>
%  <concept_significance>300</concept_significance>
%  <concept_id>10010520.10010553.10010554</concept_id>
%  <concept_desc>Computer systems organization~Robotics</concept_desc>
%  <concept_significance>100</concept_significance>
%  <concept_id>10003033.10003083.10003095</concept_id>
%  <concept_desc>Networks~Network reliability</concept_desc>
%  <concept_significance>100</concept_significance>
%</ccs2012>
%\end{CCSXML}
%
%\ccsdesc[500]{Computer systems organization~Embedded systems}


%
%  Use this command to print the description
%
\printccsdesc

% We no longer use \terms command


\section{Introduction}
The rapid and pervasive use of social media in multiple platforms enables accessing and sharing information ubiquitously. 
%The real time nature of tweeting makes it an effective communication tool that enables detection of current trends or any trending topics. 
%In addition, the various topics of tweets posted by Twitter users across the world everyday cover almost everything from daily routine to the important events such as sports and politics.   
In recent years, Twitter has emerged as the most popular form of microblogging and a powerful medium for communication as well as the source of information particularly during natural disasters. Studies~\cite{dufty2012using} have shown that the use of Twitter has increased drastically during and immediately after natural disasters, and it enhances the situation awareness \cite{yin2012using} via transmitting vital information in real-time. 
%\cite{mills2009web,starbird2010chatter,vieweg2010microblogging,yin2012using}. 
For instance, the effect of Twitter for the Great East Japan earthquake and the subsequent tsunami in the city of Tohoku in 2011 shows a strong correspondence between emergencies as mentioned in Twitter data and emergencies as handled by different organisations~\cite{kaigo2012social,doan2012analysis}. 
%\cite{kaigo2012social,doan2012analysis}. 
It is also shown that Twitter provided vital information and knowledge for the citizens of Tsukuba as a back-channel while many of the lifelines were not functioning effectively~\cite{kaigo2012social}. In another study, Starbird and Palen \cite{starbird2010pass} examine the more effective role of retweets (RT@) for information propagation in a mass emergency. 

The primary problem with using social media in critical situation such as disaster management is the presence of \textit{noise} as spam and misinformation (both intentional and unintentional). Additionally, in most natural disasters, the volume of Twitter data becomes exceptionally high. It is an overwhelming task for humans to sift through this large amount of data to find any actionable information. This highlights the need for automated approaches for analysing such data in this domain. 

Although analysing social media in the domain of disaster management has been studied in the literature
%\cite{abel2012semantics,Ashktorab2014Tweedr,Karimi:2013,kumar2011tweettracker,rogstadius2013crisistracker}
 \cite{Ashktorab2014Tweedr,rogstadius2013crisistracker}, the issues of automated and real-time monitoring and visualization of important features of large volumes of tweets have not been properly addressed. In this demo, we present Australian Crisis Tracker (ACT) which operated as a pilot during the $2013$-$2014$ bushfires season at the Australian Red Cross (ARC) - Emergency Operation Centre (EOC). Our main focus in developing ACT is to minimise the visual clutter of the high volume of data and provide the end-users with important and essential information about the unfolding event. 

\section{Australian Crisis Tracker (ACT)}\label{sec:Framework}
\begin{figure} 
\begin{center} 
\includegraphics[width=1\columnwidth, width=85mm]{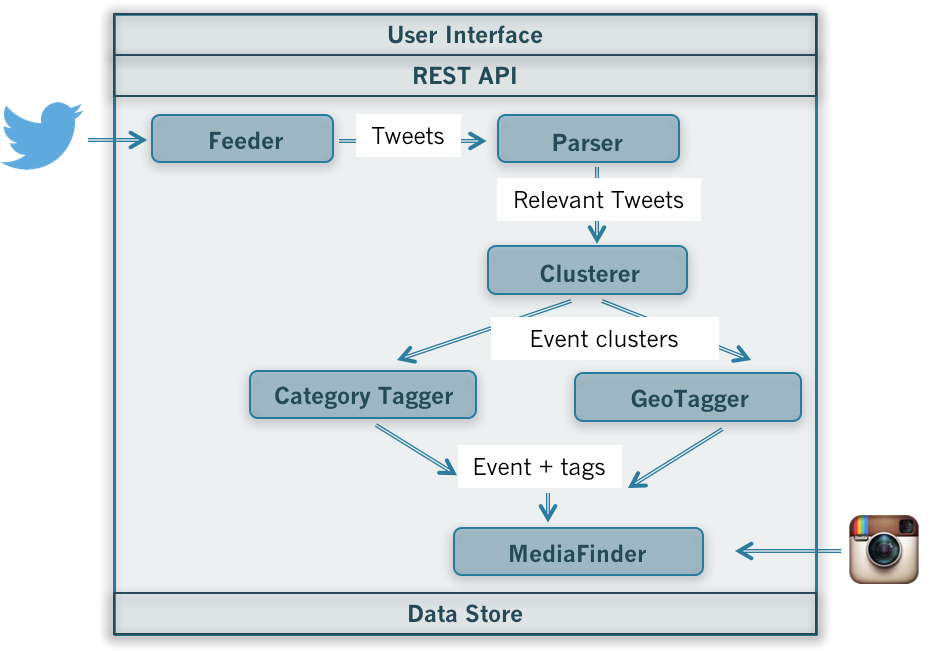} 
\caption{ACT Framework \label{fig:framework}}  
\end{center}  
\end{figure}

Figure~\ref{fig:framework} shows the ACT system for ingesting, analysing and correlating Twitter and Instagram data for disaster management. The principle source of data in ACT is the tweets collected from the Twitter Streaming API by the ``Feeder'' module. The ``Feeder'' tracks relevant Twitter accounts for disaster management such as emergency services, Non-governmental aid organisations and news media, as well as  a set of keywords associated with natural disasters, such as flood, fire, bushfire, hurricane, ambulance, etc. The "Parser" then filters out the irrelevant tweets such as spam, jokes, songs, etc., collects relevant metadata from the incoming tweets, and stores them in a database. ``Clusterer'' then groups relevant tweets which form \textit{events}. Next, the ``GeoTagger'' and the ``Category Tagger'' implement more complex methods to determine the geolocation of the event as mentioned in the Tweet and the category of the events. Finally, ``Media Finder'' searches and extracts relevant images from Instagram using the content, location and category of the events. At the end of the pipeline, the analytics is presented in the ``User Interface'' designed specifically for the ACT framework.

\subsection{Interaction and Analytics in ACT}\label{sec:UI}

\begin{figure}
\begin{center} 
\includegraphics[width=1\columnwidth]{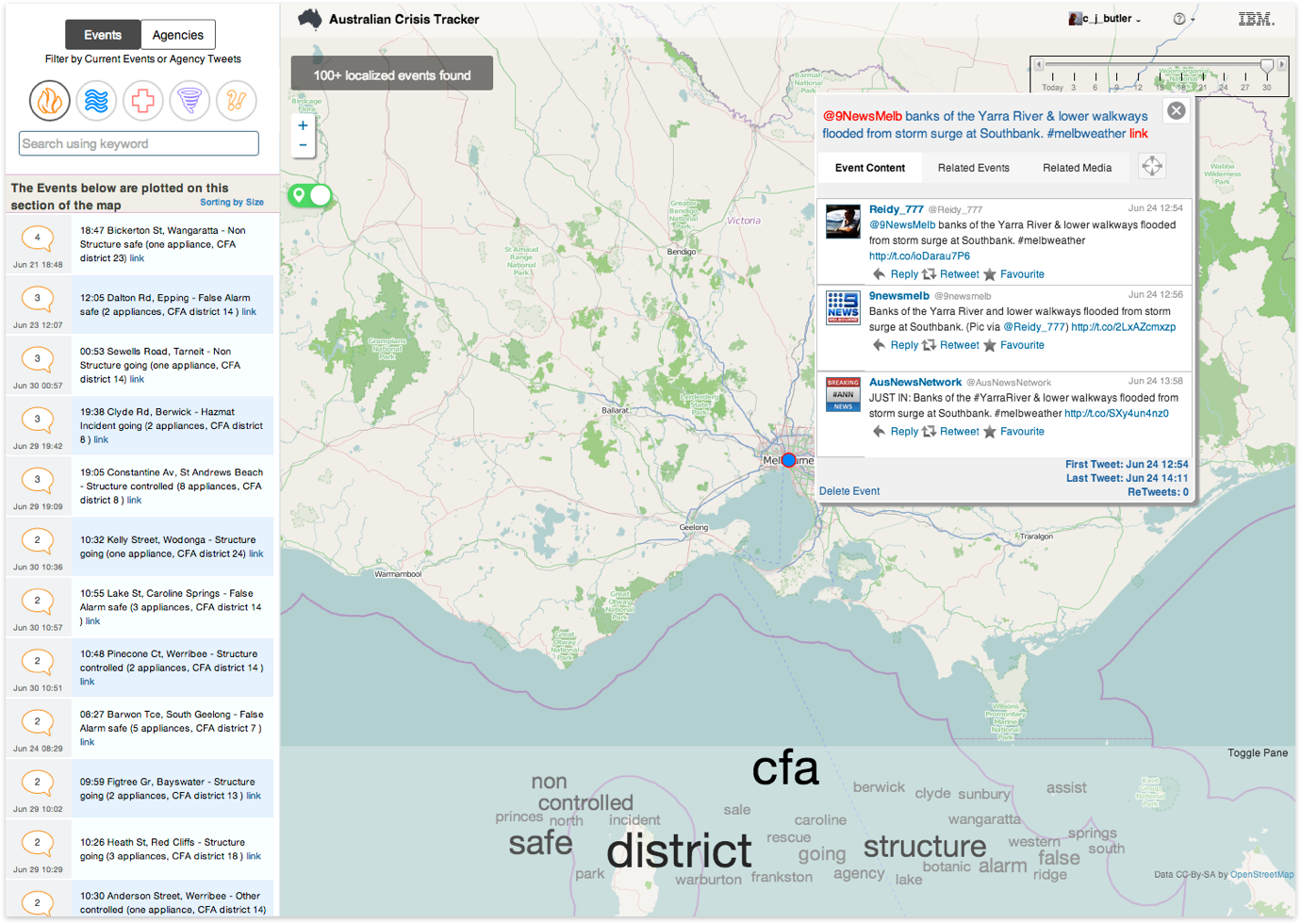}
\caption{Primary Interface of ACT}
\label{fig:UI-overview}
\end{center} 
\end{figure}

This section describes the interaction and analytics available in ACT. The ACT user interface is designed as a single web page application to
allow users to interact with the Twitter and Instagram information pertaining to events of interest to the user. As shown in Figure~\ref{fig:UI-overview}, the left hand side bar serves the purpose of displaying a set of events in a list based on the current filter criteria. Users have the choice either to explore the events or to explore the individual tweets of the known sources, such as ``Agencies''.

The event tab allows following functionalities: (i) specifying a category, (ii) specifying a keyword
search, and (iii) toggling between events with and without geo tags. The latter function is designed
to specifically consider the geolocation of the events. For example, if the events are geotagged,
the bounding box of the map acts as a geographical filter on the event data, while the events are
not geotagged, the filtering ability is restricted to time, keyword and category only. An additional
filter is also set to specify the historical range on the right hand side of the map. This allows exploration of historical events information. The  bottom of the map contains a word cloud which presents the user with a quick snapshot of trending terms pertaining to the current selection. 

In addition to the above features, clicking on either an event in the sidebar or a marker on the map
brings up an event dialogue box containing three tabs: event content (each tweet where the text content is unique for the event), related events (similar in terms of geospatial location, time of the event and categorisation) and event media (images from twitter and Instagram). 

\subsection{ARC Feedback}
Despite the relatively sparse utilisation of ACT by the ARC during the bushfire season, they provided unique feedback on the operation of ACT for disaster management. Initial versions of ACT employed a ``profanity filter'' in order to remove explicit content, but ARC recommended to capture the anger typified by expletives due to their principle interest on psycho-social aspects of disaster management. ARC also strongly welcomed the improvement to include relevant images. The initial version of ACT featured ``disaster'' icons on the map on for any event detection, but ARC perceived a risk that the map with a large number of icons could be misconceived in a time of crisis, particularly if it is distributed to the individuals not educated in the tool. The final improvement requested by the ARC was to augment ACT UI to allow its use in a ``dashboard'' fashion in addition to an interactive tool.

In response to the above suggestions, we have enhanced ACT with more features: (i) ``Sentiment analysis'' on the tweets and events for capturing anger and frustration phrases, (ii) MediaFinder module to augment events with images from Twitter and Instragram, and (iii) analytics available as APIs to enable usage of ACT in a command center ``dashboard''. 

\section{Conclusion and Future Work}
This demo presented an automated framework, named Australian Crisis Tracker (ACT), for real-time social media data analysis for natural disasters. By utilising a combination of categorical and location-based filters, the system is able to detect the tweets and images related to similar events, which provides the users with the interface to drill down to the insight rather than looking into an isolated tweet or image. 
Moreover, ACT provides an interactive visualisation platform for the users to explore the events. 
The overall goal of the system is to minimise manual effort in extracting relevant information from social media data to accelerate the required recovery actions during the natural disasters. 
We plan to extend the ACT framework to be applicable in any general domain besides natural disasters and make it accessible via both web and mobile platforms with personalised customisations.

%\end{document}  % This is where a 'short' article might terminate

%ACKNOWLEDGMENTS are optional
%\section{Acknowledgments}
%This section is optional; it is a location for you
%to acknowledge grants, funding, editing assistance and
%what have you.  In the present case, for example, the
%authors would like to thank Gerald Murray of ACM for
%his help in codifying this \textit{Author's Guide}
%and the \textbf{.cls} and \textbf{.tex} files that it describes.

\bibliographystyle{abbrv}
\bibliography{act}  % sigproc.bib is the name of the Bibliography in this case

\begin{thebibliography}{1}

\bibitem{Ashktorab2014Tweedr}
Z.~Ashktorab, C.~Brown, M.~Nandi, and A.~Culotta.
\newblock Tweedr: Mining twitter to inform disaster response.
\newblock In {\em Procs of the 11th ISCRAM Conference}, pages 354--358,
  University Park, Pennsylvania, USA, May 2014.

\bibitem{doan2012analysis}
S.~Doan, B.-K.~H. Vo, and N.~Collier.
\newblock An analysis of twitter messages in the 2011 tohoku earthquake.
\newblock In {\em Electronic Healthcare}, volume~91 of {\em Lecture Notes of
  the Institute for Computer Sciences, Social Informatics and
  Telecommunications Engineering}, pages 58--66. Springer, 2012.

\bibitem{dufty2012using}
N.~Dufty.
\newblock Using social media to build community disaster resilience.
\newblock {\em The Australian Journal of Emergency Management}, 27(1):40--45,
  2012.

\bibitem{kaigo2012social}
M.~Kaigo.
\newblock {Social media usage during disasters and social capital: Twitter and
  the Great East Japan earthquake}.
\newblock {\em Keio Communication Review}, 34:19--35, 2012.

\bibitem{rogstadius2013crisistracker}
J.~Rogstadius, M.~Vukovic, C.~Teixeira, V.~Kostakos, E.~Karapanos, and
  J.~Laredo.
\newblock {CrisisTracker: Crowdsourced social media curation for disaster
  awareness}.
\newblock {\em IBM Journal of Research and Development}, 57(5):4:1--4:13, 2013.

\bibitem{starbird2010pass}
K.~Starbird and L.~Palen.
\newblock {Pass It On?: Retweeting in Mass Emergency}.
\newblock In {\em Procs of the 7th ISCRAM Conference}, pages 1--10, Seattle,
  WA, USA, 2-5 May 2010.

\bibitem{yin2012using}
J.~Yin, A.~Lampert, M.~Cameron, B.~Robinson, and R.~Power.
\newblock Using social media to enhance emergency situation awareness.
\newblock {\em IEEE Intelligent Systems}, 27(6):52--59, 2012.

\end{thebibliography}
% ACM needs 'a single self-contained file'!
%\balancecolumns

\end{document}